# The effect of acceleration parameter on thermal entanglement and teleportation of a two-qubit Heisenberg XXX model with Dzyaloshinski-Moriya interaction in non-inertial frames

Chuan Mo(默川),  Guo-Feng Zhang(张国锋) *

*Key Laboratory of Micro-Nano Measurement-Manipulation and Physics (Ministry of Education), School of Physics, Beihang University, Beijing 100191, China*

**Abstract :** Thermal entanglement and teleportation properties in non-inertial frames are investigated when a two-qubit Heisenberg XXX model with Dzyaloshinski-Moriya (DM) interaction is used as the quantum channel. We find that thermal entanglement and the teleportation fidelity of a teleported state decrease with the acceleration parameter or temperature, in other words, the quantum entanglement teleportation will degenerate in non-inertial frames. Furthermore, our results also show that the average fidelity may be promoted by DM interaction for the ferromagnetic case, which is opposite to the antiferromagnetic case in non-inertial frames, although hindered by acceleration parameter, and their influence can be counteracted.

## I. Introduction

Entanglement is one of the most studied notions of quantum theory and plays an important role in quantum communication. It is crucial to distinguish quantum entangled states from the separable states. We often use concurrence to describe the entanglement of multipartite systems [1]. The shared entangled qubits allow us to realize some quantum teleportation [2]. Entanglement has become a key in quantum teleportation due to its quantum nonlocality feature [3]. Now, some excellent publications have been referred to the non-inertial frames, which had attracted great interest in the past decade [4,5]. It is significant to pay attention to the multipartite entanglement in non-inertial frames. Because it combines the theory of quantum field, quantum information and general relativity [6].

* Corresponding author. gf1978zhang@buaa.edu.cn

Recently, researches have been devoted to quantum information in non-inertial frames, which has become an interesting topic. Stimulated by these studies, we have investigated the effect of acceleration parameter on thermal entanglement and teleportation of a two-qubit Heisenberg XXX model with DM interaction in non-inertial frames. A two-qubit Heisenberg XXX model is considered as an appropriate candidate for entanglement. Moreover, T. Moriya [7] pointed out that the energy contribution of DM interaction is actually an antisymmetric part of the bilinear spin-spin interaction and it can be considered the influence of spin-orbit coupling. Thermal entangled states of Heisenberg spin chain and a two-qubit state in non-inertial frames can be used for realizing the entanglement teleportation.

In addition, it is very important to protect entanglement and teleportation, which are easily affected by the environment in real quantum information processing [8]. For example, temperature has great effect on thermal entanglement. Through the analysis of acceleration parameter and other variables, this paper is useful for qualitative and quantitative description of entanglement and teleportation properties in non-inertial frames. It will provide some guidance into the theoretical and practical levels of quantum teleportation creation and quantum communication development.

The paper is structured as follows. A two-qubit Heisenberg XXX model in the presence of DM interaction is introduced in Sec. II. Entanglement teleportation by the thermal mixed states in non-inertial frames is analyzed in Sec. III. And Sec. IV is devoted to investigate theoretically the fidelity. Finally, the conclusions are presented in Sec. V.

## II. Heisenberg XXX Model

The Heisenberg XXX model with DM interaction [9] can be given by

$$H = \frac{J}{2}[\sigma_{1x}\sigma_{2x} + \sigma_{1y}\sigma_{2y} + \sigma_{1z}\sigma_{2z} + \vec{D}\cdot(\vec{\sigma}_1 \times \vec{\sigma}_2)], \tag{1}$$

where $J$ is the real coupling coefficient, $\vec{D}$ is the DM vector coupling and $\sigma_{1i}, \sigma_{2i} (i = x, y, z)$ signify three components of the Pauli matrix for spin 1 and spin 2, respectively. For simplicity, we can assume the DM interaction is only along $Z$ direction,

$$H = \frac{J}{2}[\sigma_{1x}\sigma_{2x} + \sigma_{1y}\sigma_{2y} + \sigma_{1z}\sigma_{2z} + D(\sigma_{1x}\sigma_{2y} - \sigma_{1y}\sigma_{2x})], \tag{2}$$

and the eigenvalues and eigenvectors are described by

$$H|00\rangle = \frac{J}{2}|00\rangle, H|\pm\rangle = (\pm J\sqrt{1+D^2} - \frac{J}{2})|\pm\rangle, \tag{3}$$

$$H|11\rangle = \frac{J}{2}|11\rangle, H|-\rangle = (-J\sqrt{1+D^2} - \frac{J}{2})|-\rangle, \tag{4}$$

with $|\pm\rangle = (|01\rangle \pm e^{i\alpha}|10\rangle)/\sqrt{2}$ and $\alpha = \arctan D$.

In the standard basis $\{|11\rangle, |10\rangle, |01\rangle, |00\rangle\}$, the density matrix $\rho(T)$ of a solid state system at thermal equilibrium [10] can be expressed as

$$\rho(T) = \frac{e^{-\beta H}}{\mathrm{Tr}(e^{-\beta H})} = \frac{1}{Z}\begin{pmatrix} e^{-\frac{\beta J}{2}} & 0 & 0 & 0 \\ 0 & \frac{1}{2}e^{\beta(J-\delta)/2}(1+e^{\beta\delta}) & \frac{1}{2}e^{i\alpha}e^{\beta(J-\delta)/2}(1-e^{\beta\delta}) & 0 \\ 0 & \frac{1}{2}e^{-i\alpha}e^{\beta(J-\delta)/2}(1-e^{\beta\delta}) & \frac{1}{2}e^{\beta(J-\delta)/2}(1+e^{\beta\delta}) & 0 \\ 0 & 0 & 0 & e^{-\frac{\beta J}{2}} \end{pmatrix}, \tag{5}$$

where $Z = 2e^{-\frac{\beta J}{2}}[1 + e^{\beta J}\cosh\frac{\beta\delta}{2}]$, $\delta = 2J\sqrt{1+D^2}$ and $\beta = \frac{1}{kT}$, in which $k$ is Boltzmann constant and $T$ represents temperature. Conveniently, the Boltzmann constant can be written as $k = 1$ in the following calculation.

## III. Entanglement Teleportation in Non-inertial Frames

Entanglement teleportation can transmit quantum information from one place to another. The information transmission of entanglement teleportation can be achieved by the thermal mixed states of Heisenberg spin chain and an arbitrary pure state. As we know, a two-qubit Heisenberg XXX model with DM interaction can be used as the quantum channel. And the state in inertial frames can be given by

$$|\phi\rangle = \cos\frac{\theta}{2}|10\rangle + e^{i\varphi}\sin\frac{\theta}{2}|01\rangle, \quad (0 \le \theta \le \pi, 0 \le \varphi \le 2\pi), \tag{6}$$

where $\theta$ describes the amplitude and $\varphi$ stands for the phase of an arbitrary pure state.

In order to investigate the thermal entanglement and teleportation properties for the two qubits system in non-inertial frames, the Heisenberg XXX model is considered as a candidate channel in inertial frames, while the input state is in non-inertial frames. We often use the Rindler coordinates to divide Minkowski space-time into two regions I and II [11]. Thus, the state in non-inertial frames is considered to be located in Region I and disconnected from another situation in Region II.

The Minkowski vacuum state [12] can be expressed in terms of a product of two-mode squeezed states of the Rindler vacuum. Therefore, the respective transformations between two coordinated systems for the fermion field can be written as

$$\begin{cases} |0_{\omega_i}\rangle_M = \cos r_i\, |0_{\omega_i}\rangle_I |0_{\omega_i}\rangle_{II} + \sin r_i\, |1_{\omega_i}\rangle_I |1_{\omega_i}\rangle_{II} \\ |1_{\omega_i}\rangle_M = |1_{\omega_i}\rangle_I |0_{\omega_i}\rangle_{II} \end{cases}, \tag{7}$$

where $\cos r_i = (e^{-2\pi\omega_i c/a_i}+1)^{-1/2}$, $c$ is the velocity of light, $a_i$ is the acceleration of the $i$-th accelerated qubit and $\omega_i$ is its frequency.

Using the single mode approximation [13], we choose the qubit of states in non-inertial frames moves in uniform acceleration, so we can write the acceleration parameter $r_i = r$ in the following calculation. Thus, the input state Eq. (6) in non-inertial frames becomes

$$|\phi_{in}\rangle = \cos\frac{\theta}{2}|1\rangle(\cos r\,|0\rangle_I|0\rangle_{II} + \sin r\,|1\rangle_I|1\rangle_{II}) + e^{i\varphi}\sin\frac{\theta}{2}|0\rangle|1\rangle_I|0\rangle_{II}. \tag{8}$$

We can get some useful information from its density matrix, which is an important concept in quantum information. The input density matrix in non-inertial frames can be expressed as

$$\rho_{in} = |\phi_{in}\rangle\langle\phi_{in}| = \begin{pmatrix} \cos^2\frac{\theta}{2}\sin^2 r & 0 & 0 & 0 \\ 0 & \cos^2\frac{\theta}{2}\cos^2 r & \frac{1}{2}\sin\theta\cos r\,e^{-i\varphi} & 0 \\ 0 & \frac{1}{2}\sin\theta\cos r\,e^{i\varphi} & \sin^2\frac{\theta}{2} & 0 \\ 0 & 0 & 0 & 0 \end{pmatrix}. \tag{9}$$

The density matrix is an X type and this allows us to calculate physical quantities easily. We can obtain the output density matrix [14] in non-inertial frames, which can be described by $\rho_{out} = \sum_{ij} p_{ij}(\sigma_i \otimes \sigma_j)\rho_{in}(\sigma_i \otimes \sigma_j)$, where $p_{ij} = \mathrm{Tr}[E^i\rho(T)]\mathrm{Tr}[E^j\rho(T)]$, $\sum_{ij} P_{ij} = 1$, and $\sigma_i(i = 0, x, y, z)$ signify the unit matrix and three components of the Pauli matrix. In addition, there are $E^0 = |\psi^-\rangle\langle\psi^-|$, $E^1 = |\Phi^-\rangle\langle\Phi^-|$, $E^2 = |\Phi^+\rangle\langle\Phi^+|$, $E^3 = |\psi^+\rangle\langle\psi^+|$, $|\psi^\pm\rangle = (1/\sqrt{2})(|01\rangle \pm |10\rangle)$ and $|\Phi^\pm\rangle = (1/\sqrt{2})(|00\rangle \pm |11\rangle)$. After a simple algebra, the output density matrix in non-inertial frames can be written as

$$\rho_{out} = \begin{pmatrix} \rho_{11} & 0 & 0 & 0 \\ 0 & \rho_{22} & \rho_{23} & 0 \\ 0 & \rho_{23}^* & \rho_{33} & 0 \\ 0 & 0 & 0 & \rho_{44} \end{pmatrix}, \tag{10}$$

with $\rho_{11} = \frac{e^{\beta(J-\delta)}(1+e^{\beta\delta})\{2e^{\beta J}(1+e^{\beta\delta})\cos^2(\frac{\theta}{2})\sin^2 r + e^{\frac{\beta\delta}{2}}[3+\cos(2r)-2\cos\theta\sin^2 r]\}}{8[1+e^{J\beta}\cosh(\frac{\beta\delta}{2})]^2}$,

$\rho_{22} = \frac{e^{\beta J}\cos^2(\frac{\theta}{2})\cosh(\frac{\beta\delta}{2})[e^{\beta J}\cos^2 r\cosh(\frac{\beta\delta}{2})+\sin^2 r]+\sin^2(\frac{\theta}{2})}{[1+e^{J\beta}\cosh(\frac{\beta\delta}{2})]^2}$, $\rho_{23} = \frac{e^{2J\beta-i\varphi}\cos r\sin\theta\cos^2\alpha\sinh^2(\frac{\beta\delta}{2})}{2[1+e^{J\beta}\cosh(\frac{\beta\delta}{2})]^2}$,

$\rho_{33} = \frac{4\cos^2 r\cos^2(\frac{\theta}{2})+e^{\beta(J-\delta)}(1+e^{\beta\delta})[2e^{\frac{\beta\delta}{2}}\cos^2(\frac{\theta}{2})\sin^2 r+e^{\beta J}(1+e^{\beta\delta})\sin^2(\frac{\theta}{2})]}{4[1+e^{J\beta}\cosh(\frac{\beta\delta}{2})]^2}$, and

$\rho_{44} = \frac{e^{-\frac{\beta\delta}{2}}\{2e^{\frac{\beta\delta}{2}}\cos^2(\frac{\theta}{2})\sin^2 r+e^{\beta J}(1+e^{\beta\delta})[\cos^2 r\cos^2(\frac{\theta}{2})+\sin^2(\frac{\theta}{2})]\}}{8[1+e^{J\beta}\cosh(\frac{\beta\delta}{2})]^2}$.

Introduced in pioneering work, we often use the concurrence to describe the degree of the entanglement. The concurrence of two qubits [15] can be given by $C(\rho) = \max[0, \lambda_1 - \lambda_2 - \lambda_3 - \lambda_4]$, where $\lambda_i (i = 1,2,3,4)$ are the square roots of the eigenvalues of the matrix $R = \rho\rho^T$, $\rho$ is the density matrix, $\lambda_1 \geq \lambda_2 \geq \lambda_3 \geq \lambda_4$, $\rho^T = S\rho^* S$ and $S = \sigma_{1y} \otimes \sigma_{2y}$. Based on calculated concurrence, we can obtain the concurrence of the input state is $C_{in} = \sin\theta \cos r$ in non-inertial frames.

This result can be seen from FIG. 1. It shows that the concurrence is symmetric with respect to the amplitude $\theta = \pi/2$. The entanglement of input state can arrive at a maximum entangled state $C_{in} = 1$ by adjusting the amplitude $\theta = \pi/2$ and acceleration parameter $r = 0$. It is obvious that the degree of entanglement first increases with the amplitude from zero to maximum and then decreases with it to be zero, but always decreases with the acceleration parameter in non-inertial frames.

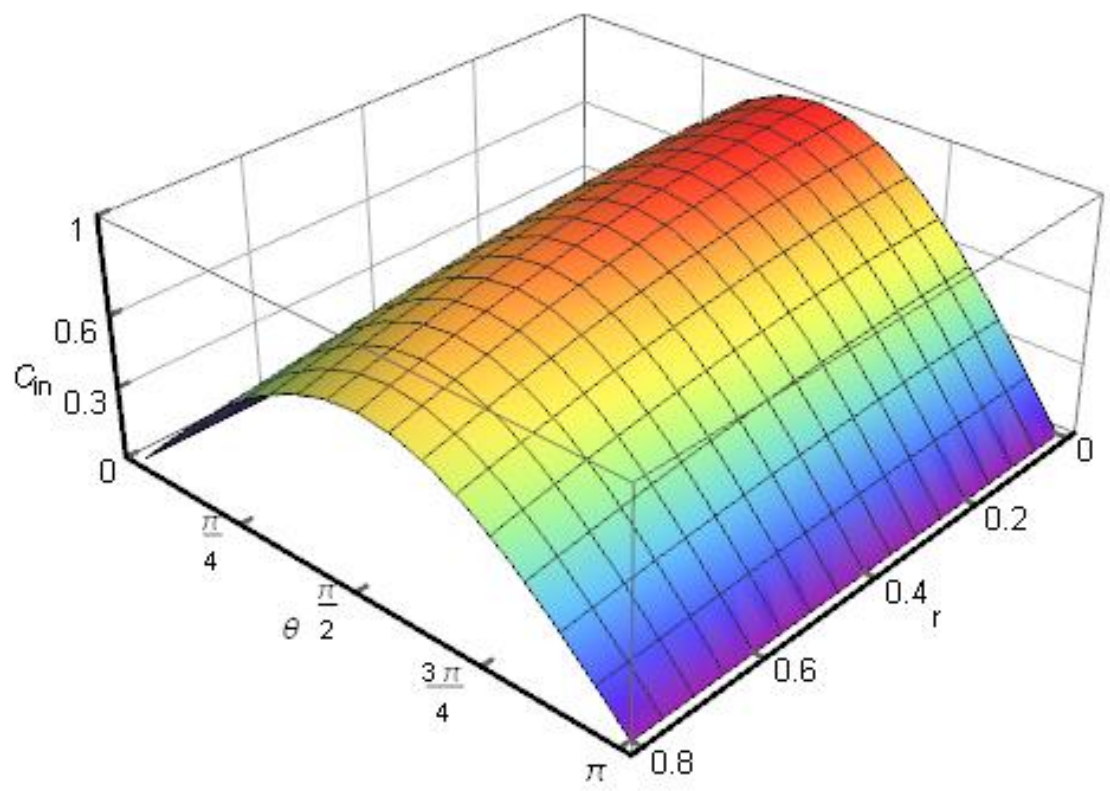

FIG. 1. Input concurrence $C_{in}$ as a function of amplitude $\theta$ and acceleration parameter $r$.

To find the characteristic of the teleported state, we also calculate the concurrence of the output density matrix $\rho_{out}$ in non-inertial frames, which can be expressed as

$$C_{out} = \max\left[\frac{e^{2J\beta} \cos r \sin\theta \cos^2\alpha \sinh^2(\frac{\beta\delta}{2})}{[1+e^{J\beta}\cosh(\frac{\beta\delta}{2})]^2} - 2\sqrt{\rho_{11}\rho_{44}}, 0\right]. \tag{11}$$

We illustrate the output entanglement in Fig. 2, and assume the temperature $T = 0.1$ considering the third law of thermodynamics, due to absolute zero is a theoretical temperature and is thought to be unreachable. It can be shown that the output entanglement of teleported state is also symmetric with respect to the amplitude $\theta = \pi/2$ and decreases with the increase of acceleration parameter, which is consistent with the input entanglement in non-inertial frames.

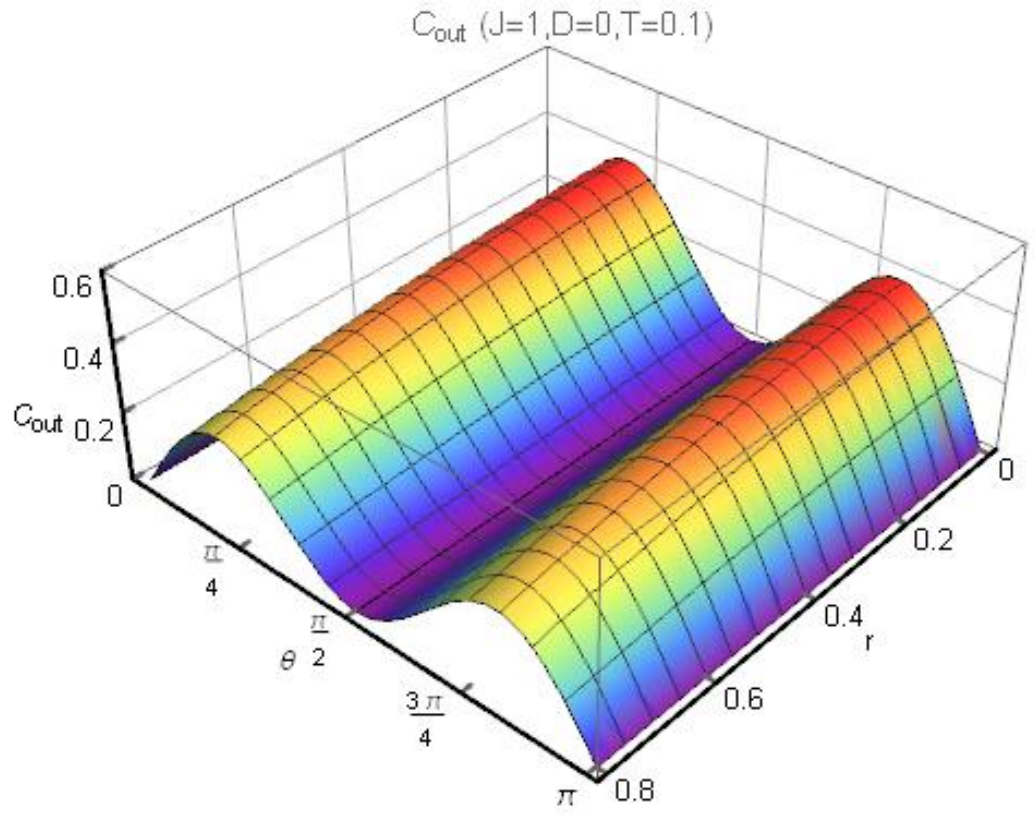

FIG. 2. Output concurrence $C_{out}$ as a function of amplitude θ and acceleration parameter r for without DM interaction and temperature T=0.1.

As can be seen from the Fig. 2, the entanglement of the teleported state is associated with the amplitude. For further discovery the relationship between the concurrence and other variables, we can take amplitude $\theta = \pi/4$ and find the effect of acceleration parameter on thermal entanglement in non-inertial frames. First, we notice that the output entanglement of the teleported state decreases with the temperature from a maximum and then falls to minimum, and it also decreases with the acceleration parameter in non-inertial frames, which corresponds to FIG. 3.

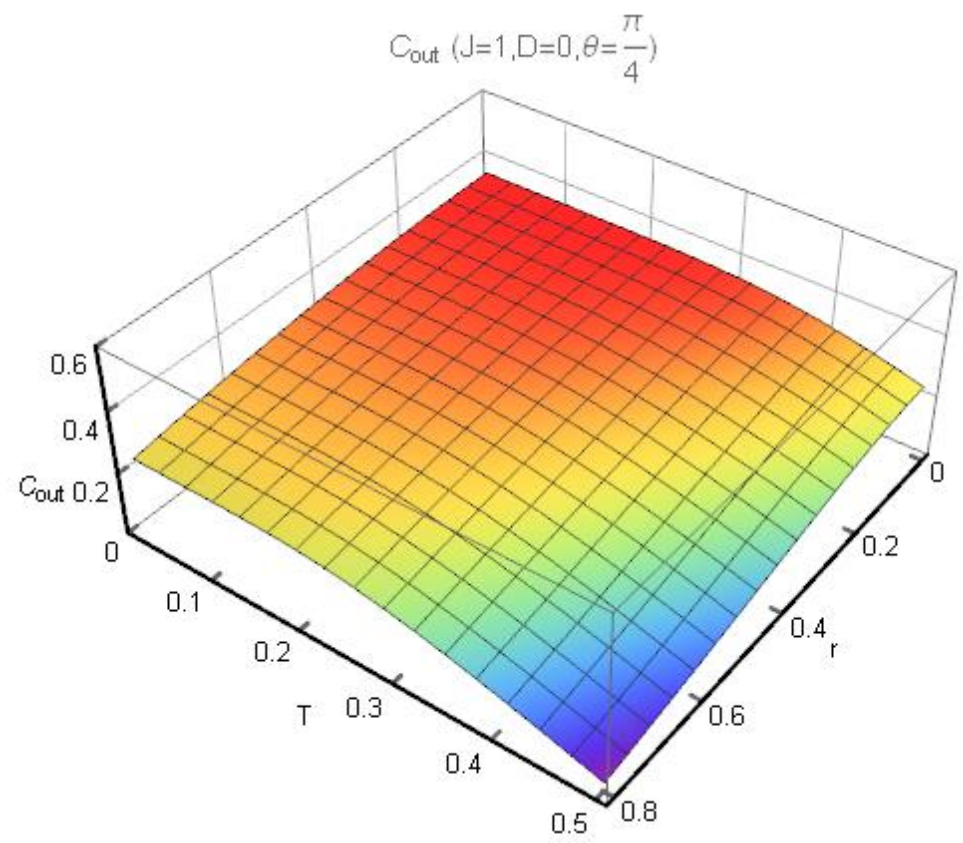

FIG. 3. Output concurrence $C_{out}$ as a function of temperature $T$ and acceleration parameter $r$ when DM interaction $D = 0$, spin coupling $J = 1$ and amplitude $\theta = \pi/4$.

Furthermore, the DM interaction has an important impact on the degree of entanglement. As we know, the real coupling coefficient $J > 0$ corresponds to the antiferromagnetic case and $J < 0$ corresponds to the ferromagnetic case. For $J > 0$, the DM interaction has little positive effect on the output entanglement based on the calculated output concurrence. However, when the real coupling $J < 0$, for example, $J = -1$ can be considered, as shown in Fig. 4. The result shows that the output entanglement increases from zero to a certain value and the DM interaction has much obviously contributory on the output entanglement in non-inertial frames.

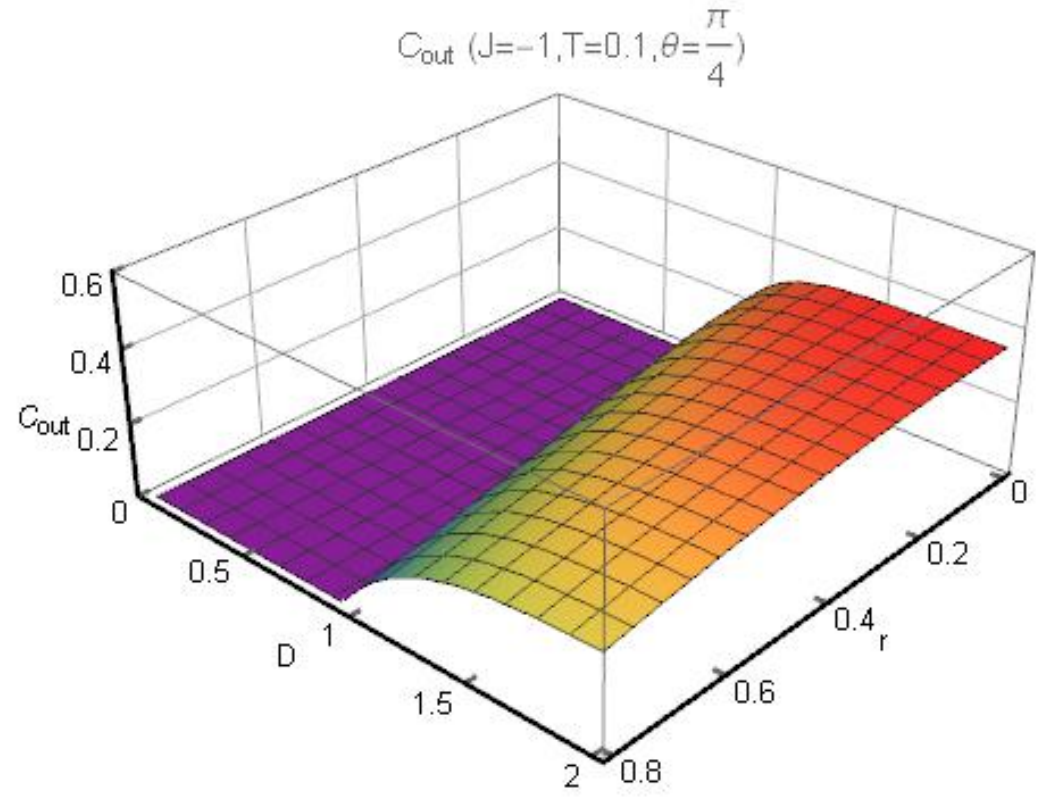

FIG. 4. Output concurrence $C_{out}$ as a function of acceleration parameter $r$ and DM interaction $D$ when temperature $T = 0.1$, spin coupling $J = -1$ and amplitude $\theta = \pi/4$.

## IV. The Average Fidelity of Entanglement Teleportation

The fidelity [16] between input density matrix $\rho_{in}$ and output density matrix $\rho_{out}$ can be formulated as $F(\rho_{in}, \rho_{out}) = \{\mathrm{Tr}\sqrt{(\rho_{in})^{1/2}\rho_{out}(\rho_{in})^{1/2}}\}^2$, which is often chosen to analysis teleportation quality. To distinguish the quality of the teleported state, we can also use the average fidelity of entanglement teleportation, which is defined by $F_A = \int_0^{2\pi} d\varphi \int_0^{\pi} F\sin\theta d\theta/4\pi$. Based on the definition, the average fidelity of our model in non-inertial frames can be written as

$$F_A = \frac{1}{48(1+D^2)[1+e^{\beta J}\cosh(\frac{\beta\delta}{2})]^2}\{e^{2J\beta}[15+11D^2+4(2+D^2)\cos(2r)]\cosh(\beta\delta)$$

$$+7e^{2J\beta}+11D^2e^{2J\beta}+4D^2e^{2J\beta}\cos(2r)+2(1+D^2)e^{2J\beta}\cos(4r)\cosh^2(\frac{\beta\delta}{2}) \quad (12)$$

$$+8e^{J\beta}(1+D^2)[2+\cos(2r)]\cosh(\frac{\beta\delta}{2})\sin^2 r+16(1+D^2)\cos^2 r\}.$$

When the effect of acceleration parameter is not considered, the entanglement teleportation is realized in inertial frames at present. Therefore, we can write the average fidelity of teleportation in non-inertial frames Eq. (12) with $r = 0$ and the average fidelity [17] in inertial frames can be given by

$$F_{A1} = \frac{2(1+\mathrm{D}^2)+e^{2\beta J}[1+2D^2+(3+2D^2)\cosh\beta\delta]}{6(1+D^2)[1+e^{\beta J}\cosh(\frac{\beta\delta}{2})]^2}. \quad (13)$$

With the influence of acceleration parameter, we notice that the average fidelity $\mathrm{F_A}$ in non-inertial frames is always smaller than $\mathrm{F_{A1}}$ in inertial system. It can also be seen from Eq. (12) and Eq. (13) or the following figures.

FIG. 5 shows the relationship of the average fidelity with temperature and acceleration parameter. The bold curve represents the average fidelity $F_{A1}$ in inertial frames without the effect of acceleration parameter $r = 0$. And the maximum value of fidelity for classical communication is $2/3$ can be seen from the axes. We notice that the average fidelity changes with the acceleration parameter without DM interaction and $J = 1$, with the development of the acceleration parameter, the average fidelity of accelerated qubit becomes smaller. When the temperature approaches zero, the average fidelity is at the maximum value, and decreases with the increase of temperature, which is consistent with the entanglement properties in non-inertial frames.

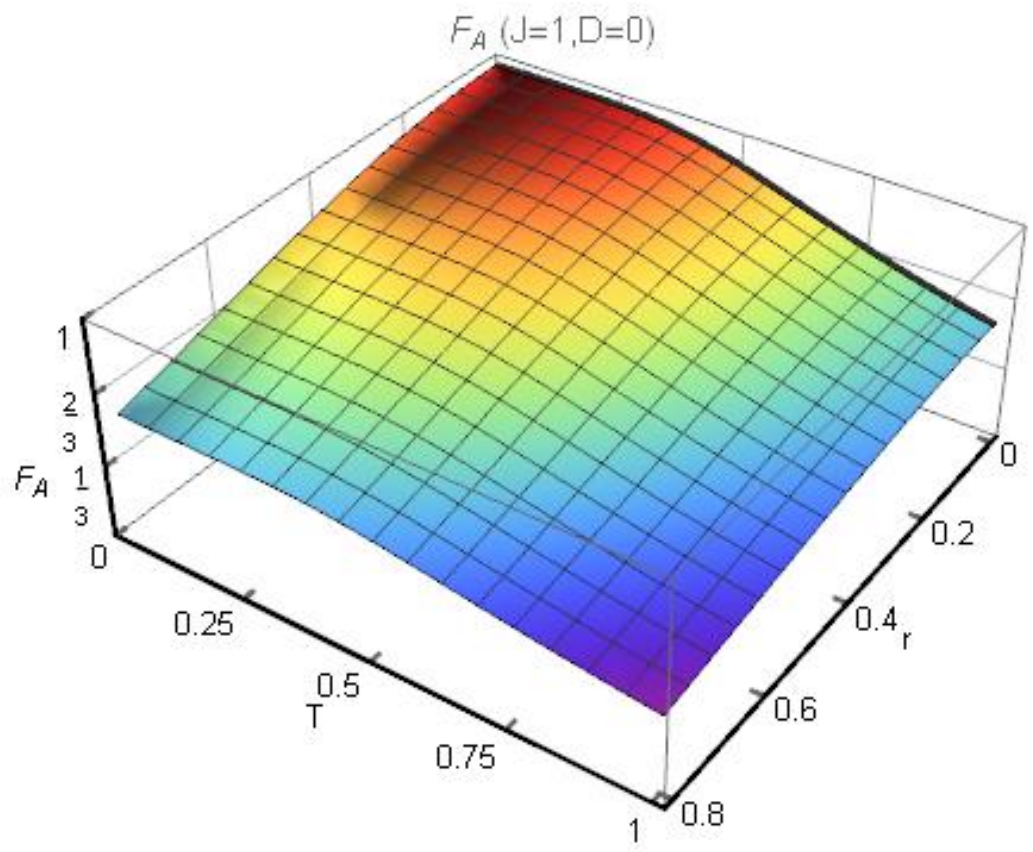

FIG. 5. Average fidelity $F_A$ as a function of acceleration parameter $r$ and temperature $T$ when DM interaction $D = 0$ and spin coupling $J = 1$.

Moreover, we also find out the effect of acceleration parameter and DM interaction and show the behavior of the average fidelity in FIG. 6. The bold curve is the same as above, representing $F_{A1}$ in inertial frames. For the antiferromagnetic case, when the real coupling coefficient $J = 1$ and $T = 0.1$, it shows that the value of average fidelity decreases with the DM interaction, and also lessens with the acceleration parameter in non-inertial frames.

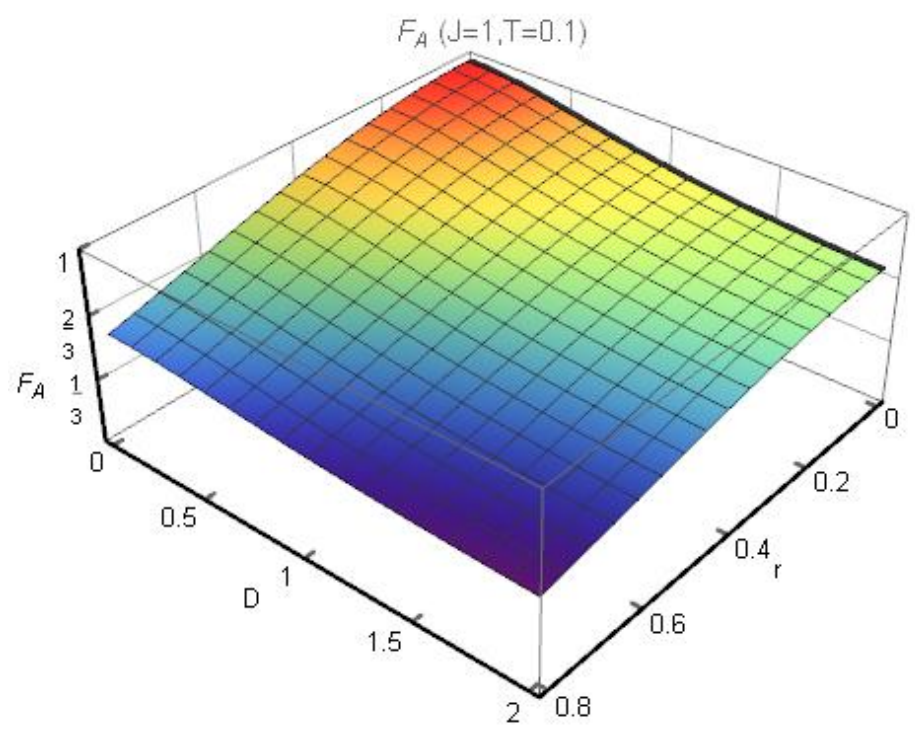

FIG. 6. Average fidelity $F_A$ as a function of acceleration parameter $r$ and DM interaction $D$ when temperature $T = 0.1$ and spin coupling $J = 1$.

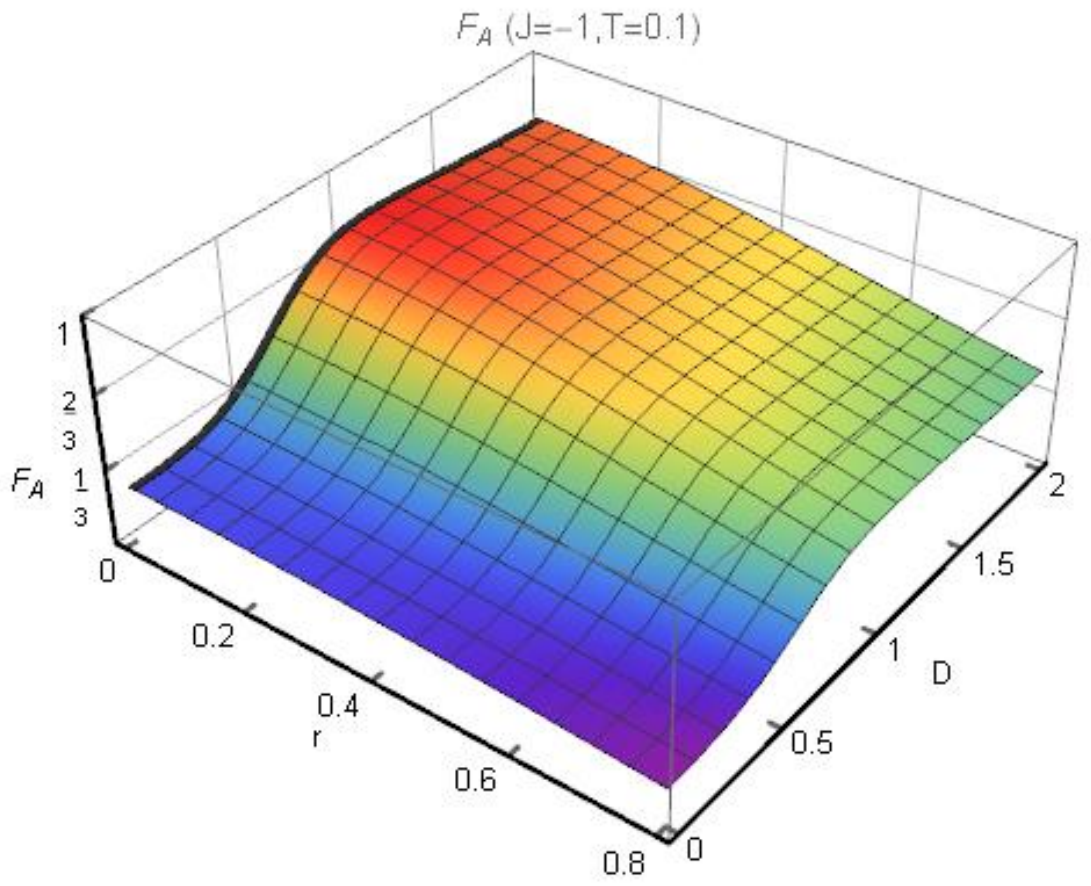

FIG. 7. Average fidelity $F_A$ as a function of acceleration parameter $r$ and DM interaction $D$ when temperature $T = 0.1$ and spin coupling $J = -1$.

However, for the ferromagnetic case, when the real coupling coefficient $J = -1$ and $T = 0.1$, FIG. 7 shows that average fidelity may be promoted by DM interaction, but decreases with the acceleration parameter. Their influence can be counteracted in non-inertial frames when the real coupling coefficient $J < 0$, which is opposite to the antiferromagnetic case.

Therefore, we notice that the average fidelity of accelerated qubit can be enhanced in non-inertial frames. We can promote the average fidelity by increasing the DM interaction for the ferromagnetic case, making up the reduction caused by the acceleration parameter. But if the qubit of state moves with an acceleration from inertial frames to non-inertial frames when other variables is the same, the average fidelity will become smaller.

Overall, the average fidelity always hindered by acceleration parameter. It is in a non-inertial frame when the accelerated qubit of the state moves with an acceleration, and the fidelity of quantum entanglement teleportation will degenerate.

These results show that we can decrease the acceleration parameter of the transmission state and reduce temperature to be a certain value to obtain the better thermal entanglement teleportation. In addition, for the ferromagnetic case, we also can enhance thermal entanglement teleportation by increasing the DM interaction when using a two-qubit Heisenberg XXX model as a quantum channel in non-inertial frames.

## V. Conclusions

We have investigated the thermal entanglement and teleportation properties of the two qubits system by thermal entangled states of a two-qubit Heisenberg XXX model with DM interaction in non-inertial frames. It is shown that amplitude and acceleration parameter affect the entanglement. Besides, the output entanglement of teleported state is related to the spin coupling, temperature and DM interaction in non-inertial frames. By characterize the quality of the teleported state, we notice that the entanglement and fidelity decrease with the increase of temperature or acceleration parameter. Moreover, the entanglement and teleportation will degenerate in non-inertial frames, so we suggest decreasing the acceleration parameter and reduce temperature to enhance the quality of teleportation when the Heisenberg XXX model is considered as a quantum channel.

## Acknowledgements

This work was supported by the National Natural Science Foundation of China (Grant No. 12074027).